\documentclass[aps,prl,twocolumn,groupedaddress]{revtex4-1}

\usepackage{lineno,graphicx, amssymb}
\usepackage{amsmath}
\usepackage{color}

\begin{document}

\setkeys{Gin}{draft=false}


\title{Coherent structures and spectral energy transfer in turbulent plasma: a space-filter approach}


\author{E. Camporeale}

\affiliation{Center for Mathematics and Computer Science (CWI), Amsterdam, The Netherlands}
\author{L. Sorriso-Valvo}
\affiliation{CNR-Nanotec - Unit\`a di Cosenza, Ponte P. Bucci, cubo 31C, 87036 Rende, Italy}
\author{F. Califano}
\affiliation{Dipartimento di Fisica "E. Fermi," Università di Pisa, Largo B. Pontecorvo 3, I-56127 Pisa, Italy}
\author{A. Retin\`{o}}
\affiliation{Centre National de la Recherche Scientifique, LPP UMR 7648, Ecole Polytechnique - Université Pierre et Marie Curie Paris VI - Observatoire de Paris, Route de Saclay Palaiseau 91128, France}


\date{\today}

\begin{abstract}
 Plasma turbulence at scales of the order of the ion inertial length is mediated by several mechanisms, including linear wave damping, magnetic reconnection, formation and dissipation of thin current sheets, stochastic heating. It is now understood that the presence of localized coherent structures enhances the dissipation channels and the kinetic features of the plasma. However, no formal way of quantifying the relationship between scale-to-scale energy transfer and the presence of spatial structures has so far been presented. In this letter we quantify such relationship analyzing the results of a two-dimensional high-resolution Hall-MHD simulation. In particular, we employ the technique of space-filtering to derive a spectral energy flux term which defines, in any point of the computational domain, the signed flux of spectral energy across a given wavenumber. The characterization of coherent structures is performed by means of a traditional two-dimensional wavelet transformation. By studying the correlation between the spectral energy flux and the wavelet amplitude, we demonstrate the strong relationship between scale-to-scale transfer and coherent structures. Furthermore, by conditioning one quantity with respect to the other, we are able for the first time to quantify the inhomogeneity of the turbulence cascade induced by topological structures in the magnetic field. Taking into account the low space-filling factor of coherent structures (i.e. they cover a small portion of space), it emerges that 80\% of the spectral energy transfer (both in the direct and inverse cascade directions) is localized in about 50\% of space, and 50\% of the energy transfer is localized in only 25\% of space.
\end{abstract}

\pacs{}

\maketitle

\section{Introduction}
Plasma turbulence has been the subject of intensive investigations, because of its importance in space, astrophysical, and laboratory applications \cite{krommes02, schekochihin09,matthaeus11,bruno13}. Amongst the several aspects that characterize plasma turbulence, such as power law exponent, spectral anisotropy , intermittency, and Alfv\'enicity \cite{sridhar94, ghosh98, sorriso01, horbury08}, much attention has recently been devoted to the role of coherent structures and their connection to turbulent dissipation and localized particle heating \cite{huld91, tenbarge13, karimabadi13, tessein13, tessein15, parashar2016, perrone17}. 
{For the purpose of the present work, by coherent structures we indicate the intermittent, spatially localized structures generated by the turbulent cascade, such as thin current sheets and magnetic eddies.}

In particular, in the context of solar wind turbulence at kinetic scale \cite{howes08a}, a somewhat dichotomous view has emerged in the community, where turbulent energy dissipation is ascribed either to linear damping of kinetic waves ---oblique propagating low-frequency Kinetic Alfv\'en Waves \cite{howes08b, sahraoui09, salem12, chen13, oughton15, coburn15} or quasi-parallel high frequency whistler waves \cite{saito08, gary09, podesta10, he11, chang11}--- or to spatially localized structures such as thin current sheets and magnetic reconnection sites \cite{osman10, perri12, wu13, osman14, cerri17}. 
Of course, both mechanisms can simultaneously be at work \cite{camporeale11, narita11, smith11, haynes14, roberts15,franci17}, however their relative importance has not yet been conclusively determined.
Several works have recently focused on studying the channels for energy transfer either in fluid or kinetic models \cite{carati06, alexakis07, aluie10, klein16, yang2017b, grete17} and in understanding the relation between dissipation enhancement and localized structures \cite{wan16, navarro16, franci17}. It is now understood, at a qualitative level, that a certain relationship between coherent structures and energy dissipation exists, but a clear assessment of such relationship is still missing.\\
In this paper we quantitatively establish the correlation between coherent structures and spectral energy transfer, analyzing a two dimensional two-fluid simulation of decaying turbulence. The spectral energy transfer is computed using a space-filter approach, a technique commonly used in Large Eddy Simulations (LES), although with a different scope (i.e. for sub-grid modeling) \cite{germano92, muller02, rivera03, ouellette12, liao13}. With the exception of a recent paper by \citet{yang17} the space-filter approach has so far been overlooked in the plasma turbulence community. {Ref. \cite{yang17} has shortly commented on the inhomogeneity of the energy flux, and the \lq coincidence between coherent structures and the sites of enhanced energy transfer\rq, without however providing a quantitative measure of such correlation.} 
In short, one can apply a filter to all variables of interest at a given wavelength and derive an equation for the conservation of filtered energy (i.e. the energy written in terms of filtered quantities), in conservative form. Such equation {contains} a source/sink term, which is of course not present in the original equation for the conservation of (unfiltered) energy. This new term has the physical meaning of spectral flux of energy across the wavelength where the filtering has been computed. The advantage of this approach, compared to the more standard global spectral decomposition, is that the spectral energy flux so derived is a quantity that is defined in the spatial domain. Its sign defines the direction of the energy cascade (towards smaller or larger scales) at a given position in space. Hence, it is straightforward to study its correlation with topological features such as spatial coherent structures. In this paper, we employ two-dimensional wavelets to derive a quantitative measure of coherent structures. We will show that the spectral energy flux and the amplitude of the wavelet transform are well correlated, indicating a larger transfer of energy (in Fourier space) in regions with strong coherent structures. Finally, by conditioning the spectral energy flux to given thresholds of wavelet amplitude, we are able for the first time to quantitatively assess the inhomogeneity of the turbulence cascade induced by topological structures in the magnetic field. In particular, taking into account the low filling-factor of coherent structures (i.e. they cover a small portion of space), it emerges that 80\% of the spectral energy transfer (both in the direct and inverse cascade directions) is localized in only about 50\% of space, 50\% of the energy transfer is localized in 25\% of space, and so on, a typical feature of intermittent turbulence \cite{frisch}.

\section{Methodology}

\begin{figure*}
 \includegraphics[width=12 cm]{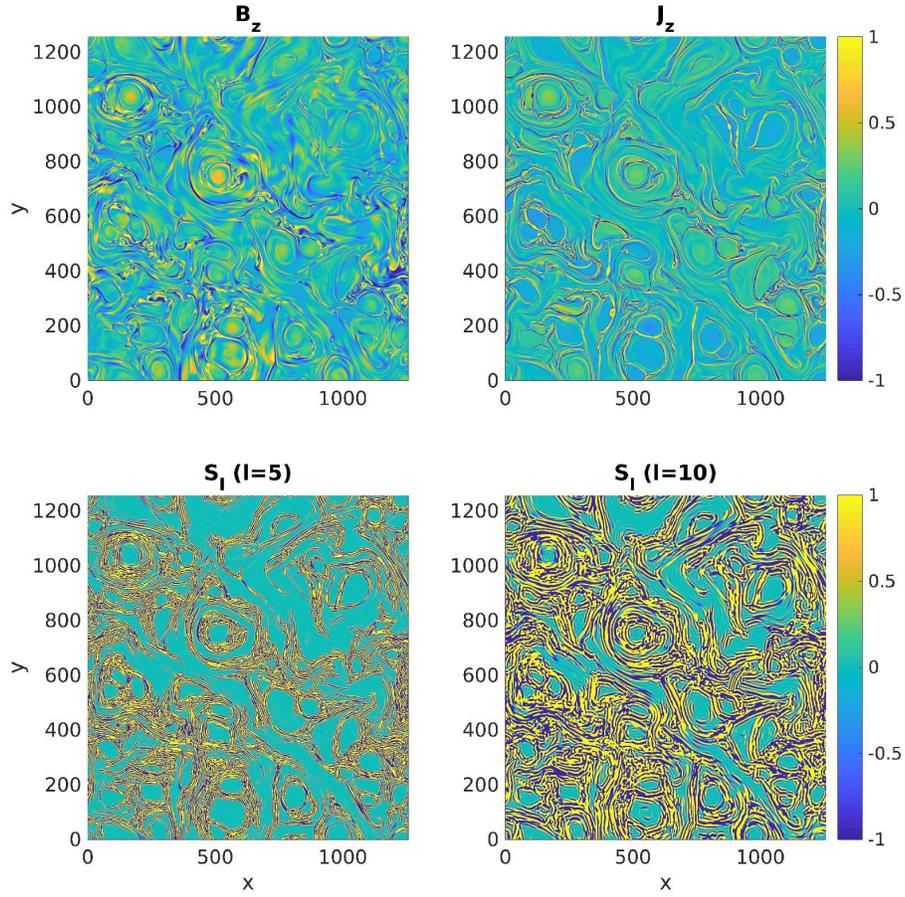}%
 \caption{Top: Snapshots of $B_z$ (left) and $J_z$ (right).  Bottom: Spectral energy flux $\mathcal{S}_l$ for $l=5$ (left) and $l=10$ (right), rescaled to the range $[-1,1]$. $x$, $y$ and $l$ are normalized to the ion inertial length.}\label{fig:1}
 \end{figure*}

Our approach is applied to a fully turbulent plasma in the two-fluids regime, i.e. Hall-MHD regime including the electron pressure gradient and electron inertia. The latter is a key ingredient to let the current sheets reconnect on a fast time scale, without dissipating the larger scales (as it would using a resistivity coefficient). The two-fluids equations are normalized to ion characteristic quantities and can be listed as the continuity and motion equation, an adiabatic closure for the pressures, the Faraday's law (neglecting the displacement current), the Ohm's law including the Hall term, the electron pressure and the electron inertia to calculate the electric field (see \cite{faganello08}). We take the mass ratio $m_e / m_i = 100$. These equations are integrated in a 2D space domain ($x, y$) of dimension $L_x = L_y = 200 \cdot 2\pi$ and using $N_x = N_y = 4096$ grid points with periodic boundary conditions. The corresponding spectrum ranges in the interval $ [ 0.005 \leq k \leq  10 ]$ where $k_{max} \simeq k_{d_e}$ (where $d_e$ is the electron inertial length).{ We impose an initial uniform out-of-plane magnetic field $B_0=1$. The initial magnetic perturbation is chosen as in \cite{cerri17b}: we excite all couples ($k_x, k_y$) laying in the semicircle $k \leq 0.015$ where $k = [k_x^2 + k_y^2]^{1/2}$ using random phases and typical mean amplitude $\epsilon \simeq 0.4$.  No initial perturbation is applied on the velocity field.} The typical eddy-turnover time turns out to be of the order of $\tau_{_L} \sim L/u_{_L} \sim 500$, much less than the final time of the simulation, $\tau_{_{fin}} = 3500$, thus allowing to obtain a fully developed regime. Indeed, the magnetic energy spectrum shows for $0.05 \leq k \leq 1$ a well-developed inertial range with spectral exponent $\alpha = 5/3$. \\
Figure \ref{fig:1} shows the snapshots of the out-of-plane magnetic field $B_z$ (top-left) and current density $J_z$ (top-right), at the time when the analysis is performed and the turbulence is well-developed. One can notice the typical formation of thin current sheets and coherent structures.

\subsection{The space-filter approach}
As it is well known, the Hall-MHD model conserves energy. Here, we seek to derive an equation for the filtered energy.
Let us consider a vector field $\mathbf{U}(\mathbf{x},t)$. We define the filtered field $\widetilde{\mathbf{U}}(\mathbf{x},t)$ via  convolution with a filter $G$ as $\widetilde{\mathbf{U}}(\mathbf{x},t)= \int_\Omega G(x-\xi)  \mathbf{U}(\mathbf{\xi},t) d\xi$
with $\Omega$ defining the entire domain.\\
The convolution can be interpreted as a low pass-filter that decomposes the field into high-frequency and low-frequency parts. In this work
we employ the so-called Butterworth filter that, in Fourier space, is $G^k = 1/\left[1+\left(\frac{k}{k_{cut}}\right)^8\right]$, where $k_{cut}$ is the wavenumber at which the filtering takes place.
Let us also introduce the Favre filter \cite{speziale88}: $ \widehat{\Phi}={\widetilde{\rho\Phi}}/{\tilde{\rho}}$, with $\rho$ the charge density.
{Obviously, the filtering of the product of two quantities is not equal to the product of the two filtered quantities (i.e., $\widetilde{\mathbf{UU}} \neq \widetilde{\mathbf{U}}\widetilde{\mathbf{U}}$). However, one can introduce so-called sub-grid residuals, that are simply defined as the difference between the two. For instance, defining  
 $\mathcal{T}_{VV} = \tilde{\rho}\widehat{VV} - \tilde{\rho}\widehat{V}\widehat{V}$, one can obtain the \emph{filtered} momentum equation $\frac{\partial \tilde{\rho}\widehat{\mathbf{V}}}{\partial t} + \nabla\cdot(\tilde{\rho} \widehat{\mathbf{V}}\widehat{\mathbf{V}}) = -\nabla\cdot(\widetilde{\mathbf{\Pi}}+ \mathcal{T}_{VV})$, and the corresponding kinetic energy equation:
$ \frac{\partial \mathcal{\widehat{E}}_U}{\partial t} =  -\nabla\cdot(\mathcal{\widehat{E}}_U\widehat{\mathbf{V}})- \nabla\cdot({\widetilde{\mathbf{\Pi}}+ \mathcal{T}_{VV}})\cdot\widehat{\mathbf{V}}
$,
where $\mathcal{\widehat{E}}_U = \frac{1}{2}\tilde{\rho} \widehat{\mathbf{V}}\cdot\widehat{\mathbf{V}} = \frac{1}{2}\tilde{\rho}(\widehat{V})^2$.
 The latter differs from the standard (unfiltered) kinetic energy equation by virtue of the sub-grid term.}
By using a combination of filtered quantities, {the same procedure can be applied to derive } an equation for the {total} \emph{filtered} energy $\widehat{\mathcal{E}}$ in conservative form (the mathematical derivation can be found in Supplemental Material, that includes Ref. \cite{cerri13}):
\begin{eqnarray}\label{energy_filtered}
 \nonumber\frac{\partial \widehat{\mathcal{E}}}{\partial t} +\nabla\cdot\left[\widehat{\mathbf{E}}\times\widehat{\mathbf{B}} +\widehat{\mathcal{E}_U}\widehat{\mathbf{V}}+\widehat{\mathcal{E}_{\Pi,i}\mathbf{V}} \right. + \\ 
 \nonumber\left.\widehat{\mathcal{E}_{\Pi,e}\left(\mathbf{V}-\frac{\mathbf{J}}{\tilde\rho}\right)} + (\widetilde{\mathbf{\Pi}_i^0}+\widetilde{\mathbf{\Pi}_i^1})\cdot\widehat{\mathbf{V}} +\widetilde{\mathbf{\Pi}_e^0}\cdot\left(\widehat{\mathbf{V}}-\frac{\widehat{\mathbf{J}}}{\tilde\rho}\right)\right] =\mathcal{S}_l
 \end{eqnarray}

The right hand side term represents the source/sink term that determines cross-scale energy transfer, with units of energy per time:
\begin{eqnarray}
 \nonumber\mathcal{S}_l = -(\nabla\cdot\mathcal{T}_{VV})\cdot\widehat{\mathbf{V}}+\left(\mathcal{T}_{V\times B}+\mathcal{T}_{J\times B}\right)\cdot\widehat{\mathbf{J}}+\\
 -\left(\mathcal{T}_{\Pi_i^0\nabla V} + \mathcal{T}_{\Pi_i^1\nabla V} +\mathcal{T}_{\Pi_e^0\nabla V} +\mathcal{T}_{\Pi_e J}\right)
\end{eqnarray}
with the following definitions
\begin{eqnarray}
 \mathcal{T}_{VV} &=& \tilde{\rho}\widehat{VV} - \tilde{\rho}\widehat{V}\widehat{V} \\
 \mathcal{T}_{V\times B} &=& \widehat{\mathbf{V}\times\mathbf{B}} - \widehat{\mathbf{V}}\times\widehat{\mathbf{B}}  \\
 \mathcal{T}_{J\times B} &=& \frac{1}{\tilde{\rho}}\left(\widetilde{\mathbf{J}\times\mathbf{B}} - \widehat{\mathbf{J}}\times\widehat{\mathbf{B}}\right)\\
 \mathcal{T}_{\Pi_s^n\nabla V} &=& \widehat{\mathbf{\Pi}_{s,kj}^n \partial_j V_k} - {\widetilde{\mathbf{\Pi}_{s,kj}^n} \partial_j \widehat{V_k}} \\
\mathcal{T}_{\Pi_e J} &=&  \widehat{\mathbf{\Pi}_{e,kj}^{0}  \partial_j \frac{J_k}{\rho}} - \widetilde{\mathbf{\Pi}_{e,kj}^{0}}  \partial_j \frac{\widehat{J_k}}{\tilde{\rho}} 
\end{eqnarray}
{Eq.(2) derives from the kinetic energy equation, Eqs.(3-4) derive from the filtering of the Hall term in Ohm's law, and Eqs. (5-6) derive from filtering the pressure equations.}
The subscript $l$ indicates the filter wavelength, that is $l=2\pi/k_{cut}$. 
The strength of the space-filter approach is that $\mathcal{S}_l$ is defined on the physical domain $(x,y)$: it is a scalar field that indicates, in each point of the domain, how much energy is transferred at a given wavelength $l$, and whose sign indicates the direction of the transfer (i.e. towards smaller or larger scales). {Contrary to the standard LES methodology, our sub-grid quantities and hence the term $\mathcal{S}_l$, can be directly calculated from simulation results.}
Two examples of $\mathcal{S}_l$ for $l=5$ and $l=10$ are shown in the bottom panels of Figure \ref{fig:1}. Here, values are normalized to the maximum value in the domain, so that the ranges are rescaled to $[-1,1]$. One can immediately identify a correlation of the spectral energy transfer with the coherent structures present in $B_z$ and $J_z$ in the top panels, similar to the results reported in \cite{yang17}.
The precise quantification of such correlation is the objective of this work.

\subsection{Coherent structures identification via wavelets}

\begin{figure}
 \includegraphics[width=12 cm]{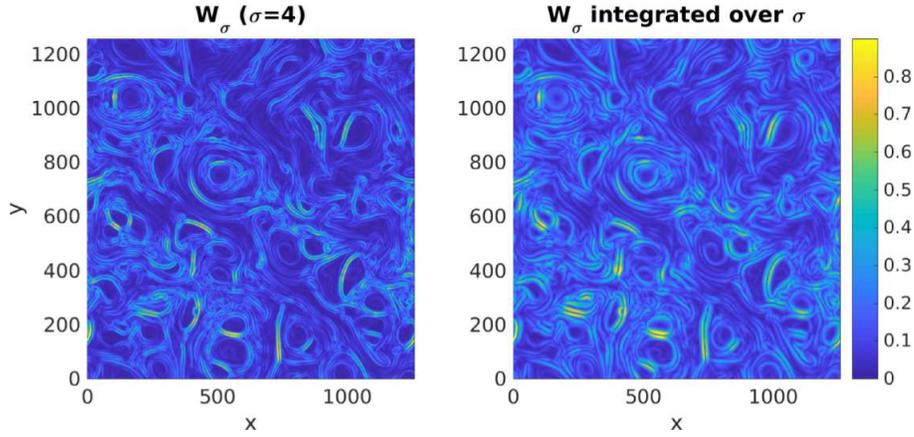}%
 \caption{Left: Wavelet amplitude $W_\sigma$ for length scale $\sigma=4$. Right: Wavelet amplitude integrated over $\sigma$. In both cases the amplitude is rescaled in the range $[0,1]$. $x$, $y$ and $\sigma$ are normalized to the ion inertial length. }\label{fig:2}
 \end{figure}

In turbulent flows, intermittency is related to the inhomogeneity of the energy cascade, which results in the appearance of small-scale energetic structures \cite{frisch}. 
The most common way to identify such structures in a $d$-dimensional field is by using the amplitude of the scale-dependent wavelet coefficients ${W}_\mathbf{\sigma}(\mathbf{r})$, where $\mathbf{\sigma}$ is a ($d$-dimensional) scale index and $\mathbf{r}$ the generic $d$-dimensional coordinate \citep{farge,onorato}. 
For example, in the solar wind, current sheets, magnetic discontinuities, and vorticity structures are commonly observed at small scales \citep{veltri,brunoLIM}. 
Studies of solar wind measurements and numerical simulations have shown that intense small-scale current sheets are statistically associated with enhanced plasma heating and other forms of ions and electrons energization \cite{osman2012, servidio2012, tessein13, Chasapis2015}. The possible processes leading to the conversion of the energy associated with turbulent fluctuations into particle energization may include magnetic reconnection, plasma instabilities and enhancement of collisions, and are still not understood \citep{chen}.
In this work we use the isotropic {\it Mexican hat} wavelet transform applied to the magnetic field to obtain the coefficients ${W}_\sigma(x,y)$, with $\sigma_x=\sigma_y=\sigma$. {The popular Mexican hat wavelet has successfully been employed for spatial structure identification in turbulent flows \cite[e.g.,][]{li98}}.
{We first compute the wavelet transform on each component of the magnetic field, and then define the total amplitude $W_{\sigma}$ as the square root of the sum of the three components squared.}
An example of the {real-valued} amplitude of the wavelet transform at the scale $\sigma=4$ (in units of ion inertial lengths) is shown in Fig. \ref{fig:2} (left). The ability of the two-dimensional wavelet to capture coherent structures is evident. The amplitude $W_\sigma$, here normalized between 0 and 1, is modulated by the intensity (gradient) of a spatial structure. In principle one could easily study the correlation between $\mathcal{S}_l$ and $W_\sigma$, as function of both $l$ and $\sigma$. However, the dependence on $\sigma$ adds an unnecessary layer of complexity that we wish to simplify. Hence, we are interested in a quantity that does not depend on $\sigma$, still retaining the ability of quantify coherent structures. A simple choice is to integrate $W_\sigma$ over all values of $\sigma$. Numerically, this translate into calculating $W_\sigma$ for a sufficiently large number of $\sigma$ and to carefully check that the integral does not depend on the choice of the range and the discretization of $\sigma$. The result, which we simply call $W$, is shown in the right panel of Fig. \ref{fig:2}. 

\section{Results}
\begin{figure}
 \includegraphics[width=12 cm]{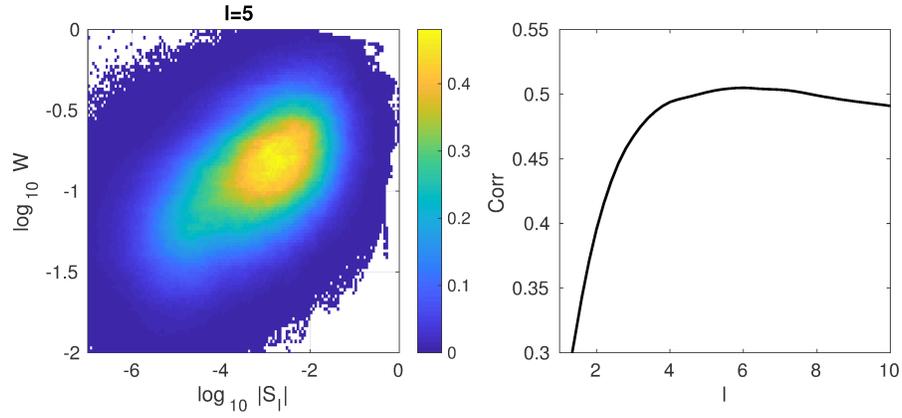}%
 \caption{Left: Colormap of the joint distribution of the quantities $\log_{10} |\mathcal{S}_l|$ and $\log_{10} W$, for $l=5$. Right: Spearman correlation coefficient between $S_l$ and $W$ ($\mathbf{J}\cdot \mathbf{E}$) in black (red), as a function of $l$. }\label{fig:3}
 \end{figure}

\begin{figure}
 \includegraphics[width=12cm]{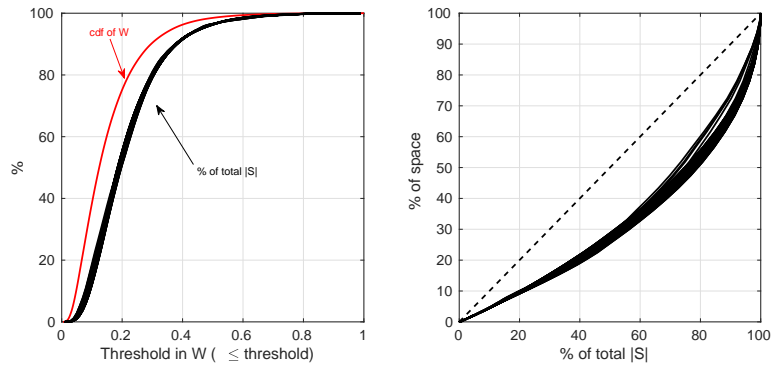}%
 \caption{Left: the red curve represents the cumulative distribution function (cdf) of the integrated wavelet amplitude $W$ over the whole computational domain. The black lines show the percentage of $|\mathcal{S}_l|$ conditioned to a given threshold in $W$, for different values of $l$. Right: each black line represents the percentage of space (vertical axis) in which a given percentage of total $|\mathcal{S}_l|$ is localized. Different curves are for different values of $l$, ranging from 1.5 to 10.} \label{fig:4}
 \end{figure}

Once we are equipped with $W$, that quantifies the location and intensity of intermittent structures, and with $\mathcal{S}_l$, that defines 
the cross-scale spectral energy transfer at wavelength $l$, it is straightforward to calculate correlations between this two quantities, and to address the question `how is the spectral energy transfer localized in space?'
The left panel of Figure \ref{fig:3} shows an example of color-map of the joint probability distribution function of the quantities $\log_{10}|\mathcal{S}_l|$ and $\log_{10} W$, for $l=5$. A strong correlation between the two quantities emerges. The right panel shows {the Spearman correlation coefficient between $W$ and $S_l$ in black and between $\mathbf{J}\cdot \mathbf{E}$ and $S_l$ in red, as a function of the scale $l$.
$\mathbf{J}\cdot \mathbf{E}$ is the (reversible) work done by the field on the particles, and it is a quantity that necessarily contains non-reversible turbulent dissipation, hence often used as a proxy for dissipation  \cite{karimabadi13, wan15, osman15, howes18}. Of course, a word of caution is needed here, when discussing the concept of dissipation in relation to Hall-MHD simulations, that do not account for all the physics needed to properly model turbulent heating. Even though this work focuses on the cross-scale energy transfer, and its correlation with spatially localized structures, the underlying implication is that an increase in large to small scales energy transfer is necessary for dissipation and heating, since they do not occur at large scales. 
The striking similarity between red and black curves in Figure 3 supports this intuitive picture and suggests that coherent structures, energy transfer and dissipation are all correlated to a certain extent. Furthermore, the observed correlation is larger for $\mathbf{J}\cdot \mathbf{E}$ than for the localized structures, suggesting that part of the energy conversion (as estimated through $\mathbf{J}\cdot \mathbf{E}$) is related to the magnitude of the energy flux, but not directly to the amplitude of the magnetic gradient. This is in agreement with recent findings in solar wind turbulence~\citep{sorriso18}}.\\ 
Obviously, coherent structures have a low space-filling factor, meaning that they are localized in a small portion of physical space, as was already evident from Figures \ref{fig:1} and \ref{fig:2}. 
{The space-filling factor can be defined by conditioning the cumulative distribution function (cdf) of $W$ on a given threshold}. The cdf is shown in the left panel of Fig. \ref{fig:4} as a red curve (where W has been normalized, as in Fig. \ref{fig:2} in the range [0,1]). For instance, only 20\% of space (i.e. numbers of grid points in the computational domain) has $W>0.2$ and only 10\% has $W>0.3$. Figure \ref{fig:4} also shows several black lines, which denote the percentage of gross energy flux transfer at a given scale, conditioned on a certain threshold in $W$. This is formally defined as $\sum_i |\mathcal{S}_l(\mathbf{x_i})|\mathcal{H}(w_t-W(\mathbf{x_i}))/\sum_i |\mathcal{S}_l(\mathbf{x_i})|$, where $\mathcal{H}$ is the Heaviside step function, and $w_t$ is a given threshold for $W$. Notice that with this choice, we consider the total amount of spectral energy transfer, i.e. the sum of its absolute value. As it is typical in turbulence, much of the cross-scale transfer cancels out, that is $\sum |\mathcal{S}_l|\gg\sum \mathcal{S}_l$.
The different black curves in the left panel of Fig. \ref{fig:4} are derived for different values of wavelength $l$ in $\mathcal{S}_l$, ranging from 1.5 to 10. Two interesting features emerge. First, the distribution of $\mathcal{S}_l$ conditioned on $W$ does not strongly depend on $l$. That is, the black curves nicely align. Second, there is a vertical gap between the red and black curves, denoting an inhomogeneity in how the spectral energy transfer is localized in space. Such inhomogeneity is not merely due to the inhomogeneity of the coherent structures (if that was the case, red and black lines would align). In other words, there is a preference of transferring spectral energy in proximity of coherent structures (defined as regions with large values of $W$). In order to quantify this preference, we plot in the right panel of Fig. \ref{fig:4} the (compliment to 100\%) values of the black lines (in horizontal axis) against the corresponding (compliment to 100\%) values of the red line (in vertical axes), for a given threshold in $W$. The interpretation is the following: the curves in Fig. \ref{fig:4} (right panel, different curves for different values of $l$) represent the percentage of total $|\mathcal{S}_l|$ active in a given percentage of space. For instance, 20\% of spectral transfer is localized in 10\% of space, 40\% in 20\% of space, 80\% in between 50\% and 60\% of space. This is the most important result of this paper, as it quantifies, for the first time, the localization of {spectral energy transfer (and possibly of turbulent dissipation)} around coherent structures in magnetized plasma.

\section{Conclusions}
A pressing topic in magnetized plasma turbulence at small scales is the relative importance played by homogeneous linear wave damping and localized dissipation due to spatial coherent structures. 
In this paper, we have quantified for the first time how much cross-scale spectral energy transfer takes place in a given portion of space, and how this correlates with the presence of coherent structures. We have used the results of a two-dimensional Hall-MHD two-fluid plasma turbulence simulation, and applied a space-filter approach to calculate, in any point of the computational domain, the spectral energy transfer $\mathcal{S}_l$ active at wavelength $l$. We have used two-dimensional isotropic Mexican hat wavelets to identify coherent structures. A clear correlation between $\mathcal{S}_l$ and the integrated wavelet amplitude $W$ emerges, with the largest correlation for $k=2\pi/l\sim 1$. By studying the cumulative distribution function of $W$ and by conditioning $S_l$ on given values of $W$, we have demonstrated that energy transfer is indeed localized in presence of strong coherent structures, which hence play a larger role in {turbulent dissipation than mechanisms mediated by linear wave damping}. However, this is not an overwhelming imbalance, but rather close to a factor of 2: 20\% of $|\mathcal{S}_l|$ is localized in about 10\% of space, 50\% in 25\% of space, and 80\% between 50\% and 60\% of space.
This is also supported by the weak increase of the correlation when the proxy for energy conversion $\mathbf{J}\cdot \mathbf{E}$ is used instead of $W$, suggesting that part of the energy conversion is still unrelated to or not co-located with the intermittent structures.
Interestingly the distribution of $|\mathcal{S}_l|$ is fairly independent from the filtering wavelength $l$.
The approach employed in this paper will be extended to kinetic simulations in future works. In this way we hope to be able to shade light to the relative importance of different kinetic mechanisms for turbulence dissipation, and their interplay with magnetic reconnection and current sheets at sub-ion scales.



%
\begin{acknowledgments}
\hfill
\\
EC was partially funded by the NWO-Vidi Grant 639.072.716.
LSV and AR acknowledge funding from the Italian CNR Short Term Mobility Programme 2014. 
The simulations were performed at CINECA (Italy) under the ISCRA initiative.
\end{acknowledgments}
\onecolumngrid

%



\newpage
\clearpage
%

\end{document}